\documentclass[a4paper,fleqn,usenatbib,onecolumn]{mnras}

\usepackage{newtxtext,newtxmath}

\usepackage[T1]{fontenc}
\usepackage{ae,aecompl}

\usepackage{graphicx}	
\usepackage{amsmath}	
\usepackage{amssymb}	

\usepackage{ulem}

\usepackage{mathtools}

\usepackage{breqn}

\newcommand{\diff}{{\rm d}}
\newcommand{\deriv}[2]{\frac{\diff #1}{\diff #2}}
\newcommand{\nderiv}[3]{\frac{\diff^{#1} #2}{\diff #3^{#1}}}

\newcommand{\ee}{{\rm e}}
\newcommand{\ii}{{\rm i}}

\newcommand{\vxi}{\boldsymbol{\xi}}
\newcommand{\xir}{\xi_{r}}
\newcommand{\xit}{\xi_{\theta}}
\newcommand{\xip}{\xi_{\phi}}

\newcommand{\vOmega}{\boldsymbol{\Omega}}

\newcommand{\Hr}{\Theta}
\newcommand{\Ht}{\widehat{\Hr}}
\newcommand{\Hp}{\widetilde{\Hr}}

\newcommand{\Lg}{\mathcal{L}^{\rm g}}
\newcommand{\Lr}{\mathcal{L}^{\rm r}}
\newcommand{\Lk}{\mathcal{L}^{\rm k}}

\title[Eigenvalues of Laplace's Tidal Equations]{Improved Asymptotic Expressions for the Eigenvalues of Laplace's Tidal Equations}

\author[R. H. D. Townsend]{
R. H. D. Townsend$^{1,2}$\thanks{E-mail: townsend@astro.wisc.edu}
\\
$^{1}$Department of Astronomy, University of Wisconsin-Madison, Madison, WI 53706, USA\\
$^{2}$Kavli Institute for Theoretical Physics, University of California, Santa Barbara, CA 93106, USA
}

\date{Accepted XXX. Received YYY; in original form ZZZ}

\pubyear{2020}

\begin{document}
\label{firstpage}
\pagerange{\pageref{firstpage}--\pageref{lastpage}}
\maketitle

\begin{abstract}
Laplace's tidal equations govern the angular dependence of
oscillations in stars when uniform rotation is treated within the
so-called traditional approximation. Using a perturbation expansion
approach, I derive improved expressions for the eigenvalue associated
with these equations, valid in the asymptotic limit of large spin
parameter $q$. These expressions have a relative accuracy of order
$q^{-3}$ for gravito-inertial modes, and $q^{-1}$ for Rossby and
Kelvin modes; the corresponding absolute accuracy is of order $q^{-1}$
for all three mode types. I validate my analysis against numerical
calculations, and demonstrate how it can be applied to derive formulae
for the periods and eigenfunctions of Rossby modes.
\end{abstract}




\section{Introduction} \label{s:intro}

Laplace's tidal equations (TEs) arise in the theory of stellar
oscillations when uniform rotation is treated within the traditional
approximation of rotation (TAR). Introduced by \citet{Eckart:1960aa},
the TAR neglects the horizontal component of the rotation
angular velocity vector $\vOmega$ when evaluating the Coriolis
force. Together with the adiabatic and \citet{Cowling:1941aa}
approximations, the TAR restores the separability of the oscillation
equations in the three spherical coordinates $(r,\theta,\phi)$. The
resulting radial ($r$) equations appear the same as in the
non-rotating case except that terms $\ell(\ell+1)$, where $\ell$ is
the spherical harmonic degree, are replaced by a separation constant
$\lambda$. This constant is found as the eigenvalue of the associated
polar ($\theta$) equations, a second-order system of differential
equations and boundary conditions comprising the eponymous TEs
first formulated by \citet{Laplace:1832aa}.

Because it greatly simplifies inclusion of the Coriolis force, the TAR
is commonly adopted in studies of waves, oscillations and tides in
rotating stars
\citep[e.g.,][]{Lee:1987aa,Bildsten:1996aa,Papaloizou:1997aa,Townsend:2005ab,Bouabid:2013aa,Fuller:2014aa,Szewczuk:2017aa,Li:2019aa}. Typically,
the TEs are solved numerically using standard techniques such as
shooting, relaxation or spectral expansion \citep[see,
  respectively,][]{Bildsten:1996aa,Fuller:2014aa,Townsend:2003ab}. However,
toward large spin parameter $q \equiv 2 \Omega/\omega$, where $\Omega
\equiv |\vOmega|$ and $\omega$ is the angular oscillation frequency in
the co-rotating reference frame, the TEs approach an asymptotic limit
where they become amenable to analytic solution. Building on earlier
work in the geophysical literature
\citep[e.g.,][]{Matsuno:1966aa,Lindzen:1967aa}, \citet[][hereafter
  T03]{Townsend:2003aa} derives approximate expressions for the
eigenvalue $\lambda$ and associated eigenfunctions (known as Hough
functions) of the TEs in this limit. These expressions are useful as
initial guesses in the aforementioned numerical techniques; they
simplify creating interpolating tables for fast TAR implementations;
and they provide the basis for estimating oscillation frequencies in
rotating stars.

In this paper I reprise the T03 analysis with the twin goals of
extending the asymptotic expressions for $\lambda$ to higher order in
$q^{-1}$, and of strengthening the mathematical
rigor. Section~\ref{s:asymp} derives the new expressions, and
Section~\ref{s:valid} validates them by comparison against numerical
calculations. Section~\ref{s:summary} then summarizes and discusses
the results of the paper.


\section{Laplace's Tidal Equations} \label{s:lte}

Within the TAR and accompanying approximations discussed in TO3, the
components of the displacement perturbation $\vxi$ for a mode with
integer azimuthal order $m$ may be expressed in the co-rotating frame as
\begin{gather}
  \label{e:xi-r}
  \xir = Y_{r}(r) \, \Hr(\theta) \, \ee^{\ii(m \phi - \omega t)}, \\
  \label{e:xi-t}
  \xit = \frac{Y_{\perp}(r)}{\sin\theta} \, \Ht(\theta) \, \ee^{\ii(m \phi - \omega t)}, \\
  \label{e:xi-p}
  \xip = \frac{Y_{\perp}(r)}{\ii \sin\theta} \, \Hp(\theta) \, \ee^{\ii(m \phi - \omega t)}.
\end{gather}
Here, $Y_{r}$ and $Y_{\perp}$ are found by solving the radial parts of
the oscillation equations (see 13--16 of T03). The Hough functions
$\Hr$, $\Ht$, $\Hp$ are likewise obtained by solving the tidal
equations
\begin{gather}
\label{e:lte-f-1}
\left[ \left(1 - \mu^{2}\right) \deriv{}{\mu} + m q \mu \right] \Hr = \left( q^{2}\mu^{2} - 1 \right) \Ht, \\
\label{e:lte-f-2}
\left[ \left(1 - \mu^{2}\right) \deriv{}{\mu} - m q \mu \right] \Ht = \left[ \lambda \left( 1 - \mu^{2} \right) - m^{2} \right] \Hr, \\
\label{e:lte-f-3}
\Hp = m \Hr - q \mu \Ht,
\end{gather}
where $\mu \equiv \cos\theta$. To avoid unphysical displacement
perturbations, the Hough functions for non-axisymmetric modes ($m \neq
0$) must vanish at the poles ($\mu = \pm 1$), and similarly for the
$\mu$ derivatives of the Hough functions for axisymmetric modes ($m =
0$).  Note that TEs~(\ref{e:lte-f-1}--\ref{e:lte-f-3}) appear slightly
different from the presentation in TO3, due to my sign choice in the
exponential terms of equations~(\ref{e:xi-r}--\ref{e:xi-p}); this
choice means that modes with $m > 0$ ($m < 0$) propagate in the
prograde (resp. retrograde) direction in the co-rotating frame.

Eliminating $\Ht$ between equations~(\ref{e:lte-f-1})
and~(\ref{e:lte-f-2}) leads to a second-order form for the TEs,
\begin{equation}
\label{e:lte-sec-r}
\deriv{}{\mu}  \left[
  \frac{1-\mu^{2}}{1 - q^{2}\mu^2} \deriv{\Hr}{\mu}
  \right] + \left\{
  m q \frac{1 + q^{2} \mu^{2}}{\left(1 - q^{2}\mu^{2}\right)^{2}} + \lambda - \frac{m^{2}}{\left( 1 - \mu^{2} \right) \left( 1 - q^{2}\mu^{2} \right)}
\right\} \Hr = 0;
\end{equation}
this is equivalent to the presentation by \citet{Bildsten:1996aa}, who
were the first explicitly to invoke the TEs in a stellar oscillation
context. An alternative second-order form can be obtained by instead
eliminating $\Hr$, yielding
\begin{equation}
\label{e:lte-sec-t}
\deriv{}{\mu} \left[
  \frac{1 - \mu^{2}}{\lambda \left(1 - \mu^{2}\right) - m^{2}} \deriv{\Ht}{\mu}
    \right] - \left\{
  m q \frac{\lambda \left(1 + \mu^{2}\right) - m^{2}}{\left[ \lambda \left(1 - \mu^{2} \right) - m^{2} \right]^{2}} +
  \frac{\lambda q^{2} \mu^{2}}{\lambda \left(1 - \mu^{2} \right) - m^{2}} - \frac{1}{1 - \mu^{2}}
  \right\} \Ht = 0.
\end{equation}
This latter form provides the starting point for the asymptotic
expressions I derive in the following section.


\section{Asymptotic Expressions} \label{s:asymp}

In the limit $|q| \rightarrow \infty$, solutions to the TEs can be
classified according to the behavior of the eigenvalue $\lambda$:
\begin{enumerate}
\item For gravito-inertial (g-i) modes, $\lambda \propto q^{2}$
\item For Rossby (r) modes, which are retrograde ($m q < 0$), $\lambda \propto q^{0}$;
\item For Kelvin modes, which are prograde ($m q > 0$), $\lambda \propto q^{0}$.
\end{enumerate}
In the following subsections I derive asymptotic expressions for
$\lambda$ for these three mode types, in the form of power-series
expansions in $w \equiv q^{-1}$. My approach is inspired by quantum
mechanical perturbation theory; the analysis is complicated by the
non-linearity of equation~(\ref{e:lte-sec-t}) in $\lambda$; but
likewise simplified by the guaranteed non-degeneracy of the
eigenvalues \citep[see, e.g.,][]{Homer:1992aa}. For each mode type,
the expansion extends to as high an order in $w$ as appears possible
while keeping the analysis relatively simple.

\subsection{Gravito-Inertial Modes} \label{s:g-asymp}

For g-i modes, I re-parameterize the TEs to use an eigenvalue
$\alpha$ and independent variable $\sigma$, where
\begin{equation} \label{e:g-alpha-sigma}
  \alpha^{2} = \lambda \, w^{2}, \qquad
  \sigma^{2} = \frac{\alpha}{w^{2}} \, \mu^{2}.
\end{equation}
The second-order form~(\ref{e:lte-sec-t}) of the TEs becomes
\begin{equation} \label{e:g-lte}
  \deriv{}{\sigma} \left[
    \frac{\alpha - w^{2} \sigma^{2}}{\alpha^{2} - w^{2} \alpha \sigma^{2} - w^{2} m^{2}} \deriv{\Ht}{\sigma}
    \right] - \left[
  m w \frac{\alpha^{2} + w^{2} \alpha \sigma^{2} - w^{2} m^{2}}{\left[ \alpha^{2} - w^{2} \alpha \sigma^{2} - w^{2} m^{2} \right]^{2}} +
  \frac{\alpha \sigma^{2}}{\alpha^{2} - w^{2} \alpha \sigma^{2} - w^{2} m^{2}} - \frac{\alpha}{\alpha - w^{2} \sigma^{2}}
  \right] \Ht = 0
\end{equation}
I then expand $\alpha$ and $\Ht$ as power series in $w$:
\begin{equation} \label{e:power-expand}
  \alpha = \sum_{i=0}^{\infty} \alpha_{i} \, w^{i}, \qquad \Ht = \sum_{j=0}^{\infty} \Ht_{j} \, w^{j},
\end{equation}
where the sequence of coefficients $\{\alpha_{0},\alpha_{1},\ldots\}$
and functions $\{\Ht_{0},\Ht_{1},\ldots\}$ will be determined. With
these expansions, equation~(\ref{e:g-lte}) has a leading order $w^{0}$
and may be expressed as
\begin{equation} \label{e:g-lte-expand}
  \sum_{i=0}^{\infty} \sum_{j=0}^{\infty} \Lg_{i} \, \Ht_{j} \, w^{i+j} = 0,
\end{equation}
where $\{\Lg_{0}\,\Lg_{1},\ldots\}$ are a sequence of second-order
linear differential operators that depend on the $\{\alpha_{i}\}$ but
not on $w$. The first four terms in this sequence are
\begin{align}
  \label{e:g-op-0}
  \Lg_{0} =& \frac{1}{\alpha_{0}} \left[ \nderiv{2}{}{\sigma} - \sigma^{2} + \alpha_{0} \right], \\
  \label{e:g-op-1}
  \Lg_{1} =& \frac{1}{\alpha_{0}^{2}} \left[ -\alpha_{1} \left( \nderiv{2}{}{\sigma} - \sigma^{2} \right) - m \right], \\
  \label{e:g-op-2}
  \Lg_{2} =& \frac{1}{\alpha_{0}^{3}} \left[ \left( m^{2} + \alpha_{1}^{2} - \alpha_{0} \alpha_{2} \right)
             \left( \nderiv{2}{}{\sigma} - \sigma^{2} \right) + 2 m \alpha_{1} + \alpha_{0}^{2} \sigma^{2} - \alpha_{0} \sigma^{4} \right], \\
  \label{e:g-op-3}
  \Lg_{3} =& \frac{1}{\alpha_{0}^{4}} \left[ \left( 2 \alpha_{0} \alpha_{1} \alpha_{2} - 3 m^{2} \alpha_{1} -
    \alpha_{1}^{3} - \alpha_{0}^{2} \alpha_{3} \right)
    \left( \nderiv{2}{}{\sigma} - \sigma^{2} \right) - m^{3} -
            3 m \alpha_{1}^{2} + 2 m \alpha_{0} \alpha_{2} - \left( 3 m \alpha_{0} + \alpha_{0}^{2} \alpha_{1} \right) \sigma^{2} +
            2 \alpha_{0} \alpha_{1} \sigma^{4} \right].
\end{align}
Equation~(\ref{e:g-lte-expand}) holds for all possible choices of $w$,
and so the coefficient of each power of $w$ must vanish. This
condition leads to a sequence of coupled differential equations, with
the first four (labeled by their corresponding power of $w$) being
\begin{align}
  \label{e:g-diff-eq-0}
  w^{0} &: \qquad \Lg_{0} \, \Ht_{0} = 0, \\
  \label{e:g-diff-eq-1}
  w^{1} &: \qquad \Lg_{0} \, \Ht_{1} + \Lg_{1} \, \Ht_{0} = 0, \\
  \label{e:g-diff-eq-2}
  w^{2} &: \qquad \Lg_{0} \, \Ht_{2} + \Lg_{1} \, \Ht_{1} + \Lg_{2} \, \Ht_{0} = 0, \\
  \label{e:g-diff-eq-3}
  w^{3} &: \qquad \Lg_{0} \, \Ht_{3} + \Lg_{1} \, \Ht_{2} + \Lg_{2} \, \Ht_{1} + \Lg_{3} \, \Ht_{0} = 0.
\end{align}
In the following sections I solve these equations in order.


\subsubsection{$w^{0}$ Equation} \label{s:g-w0-eqn}

I write equation~(\ref{e:g-diff-eq-0}) explicitly as
\begin{equation}
  \frac{1}{\alpha_{0}} \left[ \nderiv{2}{}{\sigma} - \sigma^{2} + \alpha_{0} \right] \Ht_{0} = 0.
\end{equation}
The boundary conditions at the poles require that $\Ht_{0} \rightarrow
0$ as $\sigma \rightarrow \pm \infty$. Solutions satisfying this
constraint can be found only when
\begin{equation} \label{e:g-alpha-0}
  \alpha_{0} = 2 s + 1,
\end{equation}
for integer meridional order\footnote{Section 4 of T03 discusses the
  mappings between $s$ and other mode indices.} $s \ge 0$, and can be
written
\begin{equation}
  \Ht_{0} = c_{0}\, \psi_{s},
\end{equation}
where $c_{0}$ is an arbitrary constant and $\psi_{s}$ is a normalized
Hermite function. Appendix~\ref{a:hermite} defines these functions and
presents some identities that will prove useful in the subsequent
analysis.


\subsubsection{$w^{1}$ Equation} \label{s:g-w1-eqn}

I now use the normalized Hermite functions as a basis to expand
$\Ht_{1}$ as
\begin{equation}
  \Ht_{1} = \sum_{k=0}^{\infty} c_{1,k} \, \psi_{k},
\end{equation}
where the sequence of coefficients $\{c_{1,0},c_{1,1},\ldots\}$ will
be determined. Taking the inner product between
equation~(\ref{e:g-diff-eq-1}) and $\psi_{t}$ (for arbitrary $t$) then
yields
\begin{equation} \label{e:g-prod-eq-1}
\sum_{k=0}^{\infty} c_{1,k} \left\langle \psi_{t}, \Lg_{0} \, \psi_{k} \right\rangle +
c_{0} \left\langle \psi_{t}, \Lg_{1} \, \psi_{s} \right\rangle = 0.
\end{equation}
Using the relations presented in Appendix~\ref{a:hermite}, the inner
products appearing here evaluate as
\begin{align}
  \label{e:g-overlap-0}
  \left\langle \psi_{t}, \Lg_{0} \, \psi_{k} \right\rangle &= \frac{1}{2 s + 1} \left[ (2 s + 1) - (2 k + 1) \right] \, \delta_{t,k}, \\
  \label{e:g-overlap-1}
  \left\langle \psi_{t}, \Lg_{1} \, \psi_{k} \right\rangle &= \frac{1}{(2 s + 1)^{2}} \left[ \alpha_{1} (2 k + 1) - m \right] \, \delta_{t,k}.
\end{align}
Setting $t=s$, equation~(\ref{e:g-prod-eq-1}) solves to give
\begin{equation}
  \label{e:g-alpha-1}
  \alpha_{1} = \frac{m}{2s + 1}.
\end{equation}
Likewise, with $t \neq s$ it gives
\begin{equation}
  \label{e:g-c-1}
  c_{1,t} = 0.
\end{equation}
The coefficient $c_{1,s}$ is unconstrained and can be set to an
arbitrary value; this affects the overall normalization of $\Ht$, but
is otherwise unimportant. Therefore, I choose $c_{1,s}=0$ so that
equation~(\ref{e:g-c-1}) remains true for all $t$.


\subsubsection{$w^{2}$ Equation} \label{s:g-w2-eqn}

Proceeding as before, I expand $\Ht_{2}$ as
\begin{equation}
  \Ht_{2} = \sum_{k=0}^{\infty} c_{2,k} \, \psi_{k},
\end{equation}
Taking the inner product between equation~(\ref{e:g-diff-eq-2}) and
$\psi_{t}$ then yields
\begin{equation} \label{e:g-prod-eq-2}
\sum_{k=0}^{\infty} c_{2,k} \left\langle \psi_{t}, \Lg_{0} \, \psi_{k} \right\rangle +
\sum_{k=0}^{\infty} c_{1,k} \left\langle \psi_{t}, \Lg_{1} \, \psi_{k} \right\rangle +
c_{0} \left\langle \psi_{t}, \Lg_{2} \, \psi_{s} \right\rangle = 0.
\end{equation}
Using the relations presented in Appendix~\ref{a:hermite}, the inner
product in the third term evaluates as
\begin{dmath}
  \label{e:g-overlap-2}
  \left\langle \psi_{t}, \Lg_{2} \, \psi_{k} \right\rangle =
  - \frac{\sqrt{k (k-1) (k-2) (k-3)}}{4 (2 s + 1)^{2}}  \, \delta_{t,k-4} +
  (s - k + 1) \frac{\sqrt{k (k-1)}}{(2 s + 1)^{2}} \, \delta_{t,k-2} + 
  \frac{1}{4 (2 s + 1)^{5}} \left\{
  -1 - 6 k^{2} (2 s + 1)^{3} + 2 s \left[ - 1 + 2s \left(3 - 4m^{2} + 10s + 8s^{2} \right) \right] + 
  2k \left[ (2 s + 1)^{3}(4 s - 1) - 8 m^{2} \left( 1 + 2s + 2s^{2} \right) \right] +
  4 (2 k + 1)(2 s + 1)^{3} \, \alpha_{2}
  \right\} \, \delta_{t,k} + 
  (s - k - 1) \frac{\sqrt{(k+1) (k+2)}}{(2 s + 1)^{2}} \, \delta_{t,k+2} -
  \frac{\sqrt{(k+1) (k+2) (k+3) (k+4)}}{4(2 s + 1)^{2}} \, \delta_{t,k+4}.
\end{dmath}
Setting $t=s$, equation~(\ref{e:g-prod-eq-2}) solves to give
\begin{equation}
  \label{e:g-alpha-2}
  \alpha_{2} = \frac{1 + 2s(s+1) \left[ 1 + 8 m^{2} - 4s(s+1) \right]}{4(2s + 1)^{3}}.
\end{equation}
Likewise, with $t \neq s$ it gives
\begin{equation}
  \label{e:g-c-2}
  c_{2,t} = \frac{c_{0}}{4(2s + 1)} \left[
    \frac{\sqrt{s (s-1) (s-2) (s-3)}}{8} \, \delta_{t,s-4} -
    \sqrt{s (s-1)} \, \delta_{t,s-2} -
    \sqrt{(s+1) (s+2)} \, \delta_{t,s+2} -
    \frac{\sqrt{(s+1) (s+2) (s+3) (s+4)}}{8} \, \delta_{t,s+4}
    \right].
\end{equation}
Note that this expression is not required in the subsequent analysis;
I include it here for the sake of completeness (but see the closing
comments in Section~\ref{s:summary}). Similarly to before, the
coefficient $c_{2,s}$ is unconstrained and can be set to zero, so that
the expression remains true for all $t$.


\subsubsection{$w^{3}$ Equation} \label{s:g-w3-eqn}

Again proceeding as before, I expand $\Ht_{3}$ as
\begin{equation}
  \Ht_{3} = \sum_{k=0}^{\infty} c_{3,k} \, \psi_{k},
\end{equation}
Taking the inner product between equation~(\ref{e:g-diff-eq-3}) and
$\psi_{t}$ then yields
\begin{equation} \label{e:g-prod-eq-3}
  \sum_{k=0}^{\infty} c_{3,k} \left\langle \psi_{t}, \Lg_{0} \, \psi_{k} \right\rangle +
  \sum_{k=0}^{\infty} c_{2,k} \left\langle \psi_{t}, \Lg_{1} \, \psi_{k} \right\rangle +
  \sum_{k=0}^{\infty} c_{1,k} \left\langle \psi_{t}, \Lg_{2} \, \psi_{k} \right\rangle +
  c_{0} \left\langle \psi_{t}, \Lg_{3} \, \psi_{s} \right\rangle = 0.
\end{equation}
Using the relations presented in Appendix~\ref{a:hermite}, the inner
product in the fourth term evaluates as
\begin{dmath}
  \label{e:g-overlap-3}
  \left\langle \psi_{t}, \Lg_{3} \, \psi_{k} \right\rangle =
  m \frac{\sqrt{k (k-1) (k-2) (k-3)}}{2 (2 s + 1)^{4}}  \, \delta_{t,k-4} -
  m (4s - 2k + 3) \frac{\sqrt{k (k-1)}}{(2 s + 1)^{4}} \, \delta_{t,k-2} + 
  \frac{1}{2 (2 s + 1)^{7}} \left\{
  -m + 6 m k^{2} (2 s + 1)^{3} + 4 m k \left[ 4 m^{2} \left(1 + s + s^{2}\right) - (2 s + 1)^{2} \left(1 + 4s + 7s^{2}\right) \right] -
  12 m s - 4 m s^{2} \left[14 - 4 m^{2} (s + 2) + s \left( 29 + 24s + 4s^{2} \right) \right]
  + 2 (2k + 1) (2s + 1)^{5} \, \alpha_{3}
  \right\} \, \delta_{t,k} -
  m (4s - 2k - 1) \frac{\sqrt{(k+1) (k+2)}}{(2 s + 1)^{4}} \, \delta_{t,k+2} +
  m \frac{\sqrt{(k+1) (k+2) (k+3) (k+4)}}{2(2 s + 1)^{4}} \, \delta_{t,k+4}.
\end{dmath}
Setting $t=s$, equation~(\ref{e:g-prod-eq-3}) solves to give
\begin{equation}
  \label{e:g-alpha-3}
  \alpha_{3} = m \frac{1 + 2s (s+1) \left[ 7 - 8m^{2} + 20s (s+1) \right]}{2(2s+1)^{5}}.
\end{equation}
Likewise, with $t \neq s$ it gives
\begin{dmath}
  \label{e:g-c-3}
  c_{3,t} = - \frac{m\,c_{0}}{4(2s + 1)^{3}} \left[
    \frac{\sqrt{s (s-1) (s-2) (s-3)}}{4} \, \delta_{t,s-4} -
    (2s + 3) \sqrt{s (s-1)} \, \delta_{t,s-2} +
    (2s - 1) \sqrt{(s+1) (s+2)} \, \delta_{t,s+2} -
    \frac{\sqrt{(s+1) (s+2) (s+3) (s+4)}}{4} \, \delta_{t,s+4}
    \right].
\end{dmath}
Once again, I include this expression for the sake of
completeness. The coefficient $c_{3,s}$ is unconstrained and can be
set to zero, so that the equation remains true for all $t$.


\subsubsection{Eigenvalues for Gravito-Inertial Modes} \label{s:g-eigval}

As the final step, I combine the expressions for
$\alpha_{0},\ldots,\alpha_{3}$ given in
equations~(\ref{e:g-alpha-0},\ref{e:g-alpha-1},\ref{e:g-alpha-2},\ref{e:g-alpha-3})
with the relationship~(\ref{e:g-alpha-sigma}) between $\alpha$ and
$\lambda$, and transform from $w$ back to $q$ to obtain the asymptotic
eigenvalues for gravito-inertial modes as
\begin{equation} \label{e:g-eigval}
  \lambda = q^{2} \left[ 
    (2s + 1) +
    \frac{m}{2s + 1} q^{-1} + 
    \frac{1 + 2s(s+1) \left[ 1 + 8 m^{2} - 4s(s+1) \right]}{4(2s + 1)^{3}} q^{-2} +
    m \frac{1 + 2s (s+1) \left[ 7 - 8m^{2} + 20s (s+1) \right]}{2(2s+1)^{5}} q^{-3} +
    \mathcal{O}(q^{-4})
    \right]^{2} 
\end{equation}
This can be compared against an equivalent expression obtained by a
Taylor-series expansion of the positive root in TO3's equation~(36):
\begin{equation}
  \lambda_{\rm TO3} = q^{2} \left[
    (2s + 1) +
    \frac{m}{2s + 1} q^{-1} +
    \frac{4 m^{2} s (s + 1)}{(2 s + 1)^{3}} q^{-2} +
    \mathcal{O}(q^{-3})
    \right]^{2}
\end{equation}
(I've also corrected for the different $m$ sign convention). The two
expressions differ at the third term in brackets, indicating that the
T03 expression for $\lambda$ has a relative accuracy of order
$q^{-1}$, and an absolute accuracy of order $q$.


\subsection{Rossby Modes} \label{s:r-asymp}

For r modes, I repeat the analysis of the preceding section but now
re-parameterizing via
\begin{equation} \label{e:r-alpha-sigma}
  \alpha^{2} = \lambda, \qquad
  \sigma^{2} = \frac{\alpha}{w} \, \mu^{2}.
\end{equation}
The second-order form~(\ref{e:lte-sec-t}) of the TEs then becomes
\begin{equation} \label{e:r-lte}
  \deriv{}{\sigma} \left[
    \frac{\alpha - w \sigma^{2}}{w \alpha^{2} - w^{2} \alpha \sigma^{2} - w m^{2}} \deriv{\Ht}{\sigma}
    \right] - \left[
  m w \frac{\alpha^{2} + w \alpha \sigma^{2} - m^{2}}{[w \alpha^{2} - w^{2} \alpha \sigma^{2} - w m^{2}]^{2}} +
  \frac{\alpha \sigma^{2}}{w \alpha^{2} - w^{2} \alpha \sigma^{2} - w m^{2}} - \frac{\alpha}{\alpha - w \sigma^{2}}
  \right] \Ht = 0,
\end{equation}
With the power-series expansions~(\ref{e:power-expand}) for $\alpha$ and $\Ht$,
and under the ansatz that $\alpha_{0}^{2} \neq m^{2}$, this equation
has a leading order $w^{-1}$ and may be expressed as
\begin{equation} \label{e:r-lte-expand}
  \sum_{i=0}^{\infty} \sum_{j=0}^{\infty} \, \Lr_{i} \, \Ht_{j} w^{i+j-1} = 0.
\end{equation}
The first two terms in the sequence of differential operators
$\{\Lr_{0},\Lr_{1},\ldots\}$ are
\begin{align}
  \label{e:r-op-0}
  \Lr_{0} =& \frac{1}{\alpha_{0}^{2} - m^{2}} \left[ \alpha_{0} \left( \nderiv{2}{}{\sigma} - \sigma^{2} \right) - m \right], \\
  \label{e:r-op-1}
  \Lr_{1} =& \frac{1}{(\alpha_{0}^{2} - m^{2})^{2}} \left[
    \left( m^{2} \sigma^{2} - \alpha_{0}^{2} \alpha_{1} - m^{2} \alpha_{1} \right) \left( \nderiv{2}{}{\sigma} - \sigma^{2} \right)
    + 2 m^{2} \sigma \deriv{}{\sigma} +
    m^{4} + \alpha_{0}^{4} + 2 m \alpha_{0} \alpha_{1} - 2 m^{2} \alpha_{0}^{2} 
    - 3 m \alpha_{0} \sigma^{2} + \left( m^{2} - \alpha_{0}^{2} \right) \sigma^{4} 
    \right];
\end{align}
and the resulting sequence of coupled differential equations is now
\begin{align}
  \label{e:r-diff-eq-m1}
  w^{-1} &: \qquad \Lr_{0} \, \Ht_{0} = 0, \\
  \label{e:r-diff-eq-0}
  w^{0} &: \qquad \Lr_{0} \, \Ht_{1} + \Lr_{1} \, \Ht_{0} = 0.
\end{align}
In the following sections I solve these equations in order.


\subsubsection{$w^{-1}$ Equation} \label{s:r-w-m1-eqn}

I write equation~(\ref{e:r-diff-eq-m1}) explicitly as
\begin{equation} \label{e:r-diff-eq-m1-exp}
  \frac{1}{\alpha_{0}^{2} - m^{2}} \left[ \alpha_{0} \left( \nderiv{2}{}{\sigma} - \sigma^{2} \right) - m \right] \Ht_{0} = 0.
\end{equation}
Similarly to Section~(\ref{s:g-w0-eqn}), solutions satisfying the
boundary constraint can be found only when
\begin{equation}  \label{e:r-alpha-0}
  \alpha_{0} = - \frac{m}{2 s + 1}
\end{equation}
for integer meridional order $s \ge 1$\footnote{The $s=0$ case must be
  ruled out because it violates the ansatz $\alpha_{0}^{2} \neq
  m^{2}$.}, and can be written
\begin{equation}
  \label{e:r-Ht-0}
  \Ht_{0} = c_{0} \, \psi_{s}
\end{equation}
where $c_{0}$ is an arbitrary constant.

Equation~(\ref{e:r-alpha-0}) indicates that $\alpha$ and $m$ have
opposite signs in the limit $w \rightarrow 0$. My
definition~(\ref{e:r-alpha-sigma}) of $\sigma$ requires that $\alpha$
and $w$ (or $q$) share the same sign, because $\sigma$ would otherwise
be imaginary. It therefore follows that $m$ and $q$ must have opposite
signs for r modes: these modes are necessarily retrograde.


\subsubsection{$w^{0}$ Equation} \label{s:r-w-0-eqn}

Proceeding as before, I expand $\Ht_{1}$ as
\begin{equation}
  \label{e:r-Ht-1}
  \Ht_{1} = \sum_{k=0}^{\infty} c_{1,k} \, \psi_{k}.
\end{equation}
Taking the inner product between equation~(\ref{e:r-diff-eq-0}) and
$\psi_{t}$ then yields
\begin{equation}
  \label{e:r-prod-eq-0}
  \sum_{k=0}^{\infty} c_{1,k} \left\langle \psi_{t}, \Lr_{0} \, \psi_{k} \right\rangle +
  c_{0} \left\langle \psi_{t}, \Lr_{1} \, \psi_{s} \right\rangle = 0.
\end{equation}
The inner products appearing here evaluate as
\begin{equation}
  \label{e:r-overlap-0}
  \left\langle \psi_{t}, \Lr_{0} \, \psi_{k} \right\rangle =
  (s - k) \frac{(2s + 1)}{2 m s (s+1)} \, \delta_{t,k}
\end{equation}
\begin{dmath}
  \label{e:r-overlap-1}
  \left\langle \psi_{t}, \Lr_{1} \, \psi_{k} \right\rangle =
  (2s + 1)^{2} \frac{\sqrt{k (k-1) (k-2) (k-3)}}{16 m^{2} s (s+1)}  \, \delta_{t,k-4} +
  (2s + 1)^{2} (3s - k + 2) \frac{\sqrt{k (k-1)}}{16 m^{2} s^{2} (s+1)^{2}} \, \delta_{t,k-2} -
  \frac{1}{16 m^{2} s^{2} (s+1)^{2}} \left\{
  k (2s + 1)^{2} \left[ 2s (s - 2) - 1 \right] + 2 k^{2} (2s + 1)^{2} \left( 1 + s + s^{2} \right) -
  s^{2} \left[ 16 m^{2} (s + 1)^{2} - 3 (2s + 1)^{2} \right] - 4 (2s + 1)^{2} \left[ k + s^{2} + 2 k s (s + 1) \right] \alpha_{1}
  \right\} \, \delta_{t,k} +
  (2s + 1)^{2} (3s - k) \frac{\sqrt{(k+1) (k+2)}}{16 m^{2} s^{2} (s+1)^{2}} \, \delta_{t,k+2} +
  (2s + 1)^{2} \frac{\sqrt{(k+1) (k+2) (k+3) (k+4)}}{16 m^{2} s (s+1)} \, \delta_{t,k+4}.
\end{dmath}
Setting $t=s$, equation~(\ref{e:r-prod-eq-0}) solves to give
\begin{equation}
  \label{e:r-alpha-1}
  \alpha_{1} = -\frac{1 + 2s(s+1) \left[ 1 + 8 m^{2} - 4s(s + 1) \right]}{4(2s + 1)^{3}}.
\end{equation}
Likewise, with $t \neq s$ it gives
\begin{dmath}
  \label{e:r-c-1}
  c_{1,t} = - \frac{(2s + 1)\,c_{0}}{8m} \left[
    \frac{\sqrt{s (s-1) (s-2) (s-3)}}{4} \, \delta_{t,s-4} +
    \frac{\sqrt{s (s-1)}}{s} \, \delta_{t,s-2} -
    \frac{\sqrt{(s+1) (s+2)}}{s + 1} \, \delta_{t,s+2} -
    \frac{\sqrt{(s+1) (s+2) (s+3) (s+4)}}{4} \, \delta_{t,s+4}
    \right].
\end{dmath}
The coefficient $c_{1,s}$ is unconstrained and can be set to zero, so
that this expression remains true for all $t$.


\subsubsection{Eigenvalues for Rossby Modes} \label{s:r-eigval}

As the final step, I combine the expressions for
$\alpha_{0},\alpha_{1}$ given in
equations~(\ref{e:r-alpha-0},\ref{e:r-alpha-1})
with the relationship~(\ref{e:r-alpha-sigma}) between $\alpha$ and
$\lambda$, and transform from $w$ back to $q$ to obtain the asymptotic
eigenvalues for Rossby modes as
\begin{equation} \label{e:r-eigval}
  \lambda = \left[ - \frac{m}{2s + 1} -
    \frac{1 + 2s(s+1) \left[ 1 + 8 m^{2} - 4s(s+1) \right]}{4(2s + 1)^{3}} q^{-1} +
    \mathcal{O}(q^{-2})
    \right]^{2}.
\end{equation}
This can be compared against an equivalent expression obtained by a
Taylor-series expansion of the negative root in TO3's equation~(36):
\begin{equation}
  \lambda_{\rm TO3} = \left[ - \frac{m}{2s + 1} -
    \frac{4 m^{2} s (s + 1)}{(2 s + 1)^{3}} q^{-1} +
    \mathcal{O}(q^{-2})
    \right]^{2}
\end{equation}
(again, I've corrected for the different $m$ sign convention). The two
expressions differ at the second term in brackets, indicating that the
T03 expression for $\lambda$ has a relative and absolute accuracy of
order $q^{0}$.


\begin{figure*}
  \begin{centering}
  \includegraphics{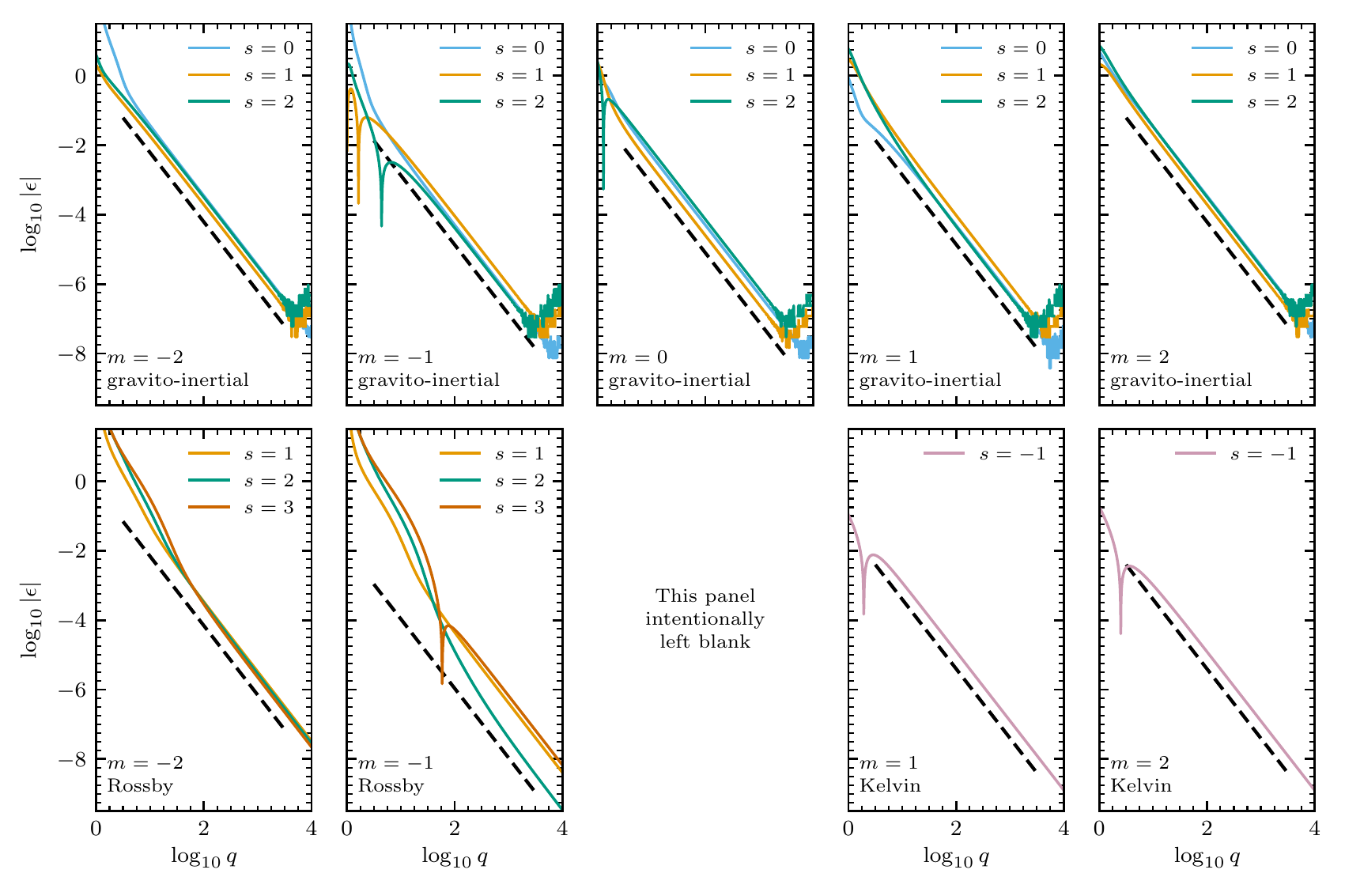}
  \caption{Log-log plots of the difference $\epsilon$ between
    asymptotic and numerical eigenvalues as a function of spin
    parameter $q$, for various azimuthal orders $m$ and meridional
    orders $s$ of each mode type. The black dashed lines indicate the
    scaling $|\epsilon| \propto q^{-2}$.} \label{f:error}
  \end{centering}
\end{figure*}

  
\subsection{Kelvin Modes} \label{s:k-asymp}

For Kelvin modes, I repeat the analysis of the preceding section but
now adopting the ansatz\footnote{A solution to
  equation~(\ref{e:r-lte}) can also be found with $\alpha_{0} = -m$;
  however, this ultimately leads to a $\Hr$ that diverges as $|\sigma|
  \rightarrow \infty$, and is therefore unphysical.} $\alpha_{0} = m$.
With the power-series expansions~(\ref{e:power-expand}) for $\alpha$
and $\Ht$, equation~(\ref{e:r-lte}) then has a leading order $w^{-2}$
and may be expressed as
\begin{equation} \label{e:k-lte-expand}
  \sum_{i=0}^{\infty} \sum_{j=0}^{\infty} \Lk_{i} \, \Ht_{j} \, w^{i+j-2} = 0.
\end{equation}
The first term in the sequence of differential operators
$\{\Lk_0,\Lk_{1},\ldots\}$ is
\begin{align}
  \label{e:k-op-0}
  \Lk_{0} =& \frac{1}{\left(\sigma^{2} - 2 \alpha_{1}^{2}\right)} \left[
    \left( 2 \alpha_{1} - \sigma^{2} \right) \left( \nderiv{2}{}{\sigma} - \sigma^{2} \right) +
    2 \sigma \deriv{}{\sigma} -
    \left( \sigma^{2} + 2 \alpha_{1} \right) \right],
\end{align}
and the first differential equation is
\begin{align}
  \label{e:k-diff-eq-0}
  w^{-2} &: \qquad \Lk_{0} \, \Ht_{0} = 0.
\end{align}
A solution to this equation satisfying the boundary constraint can
be found only when
\begin{equation} \label{e:k-alpha-1}
  \alpha_{1} = \frac{1}{4},
\end{equation}
and can be written
\begin{equation} \label{e:k-Ht0}
  \Ht_{0} = c_{0} \, \psi_{1}
\end{equation}
where $c_{0}$ is an arbitrary constant.

Combining the $\alpha_{0}=m$ ansatz with equation~(\ref{e:k-alpha-1}),
the asymptotic eigenvalues for Kelvin modes are
\begin{equation} \label{e:k-eigval}
  \lambda = \left[ m + \frac{1}{4} q^{-1} + \mathcal{O}(q^{-2})\right]^{2}.
\end{equation}
A Taylor-series expansion of TO3's equation~(55), with the usual
correction for the different $m$ sign convention, leads to the same
result, and so the latter is confirmed to have a relative and absolute
accuracy of order (at least) $q^{-1}$.

Equation~(\ref{e:k-eigval}) can also be derived by setting $s=-1$ in
the r-mode expression~(\ref{e:r-eigval}), underscoring the assignment
of a nominal meridional order $s=-1$ to Kelvin modes \citep[see,
  e.g.,][]{Gill:1982aa}. Using the same reasoning as in
Section~\ref{s:r-w-m1-eqn}, $m$ and $q$ must have the same signs for
Kelvin modes: these modes are necessarily prograde.


\begin{figure*}
  \begin{centering}
  \includegraphics{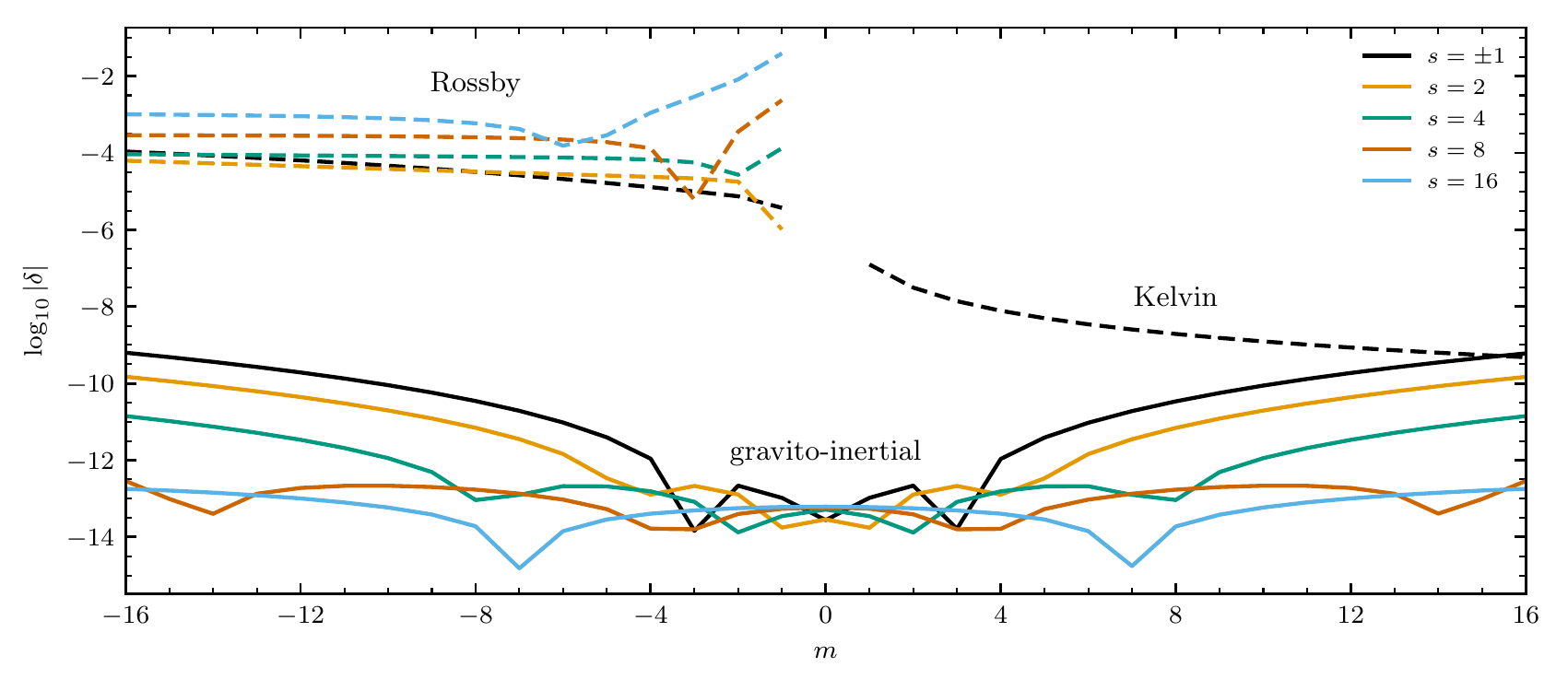}
  \caption{Log-log plots of the relative difference $\delta$
      between asymptotic and numerical eigenvalues as a function of
      spin parameter $m$, evaluated at $q=10^{3}$ and for various
      meridional orders $s$ of each mode type. Gravito-inertial modes
      are plotted with solid lines, and Rossby and Kelvin modes with
      dashed lines.} \label{f:s-m-scan}
  \end{centering}
\end{figure*}


\section{Validation} \label{s:valid}

To validate the preceding analysis, Fig.~\ref{f:error} compares the
eigenvalue
expressions~(\ref{e:g-eigval},\ref{e:r-eigval},\ref{e:k-eigval})
against numerical calculations, for azimuthal orders $-2 \leq m \leq
2$ and selected meridional orders $s$ of each mode type.  Each panel
plots $\log_{10} |\epsilon|$ as a function of $\log_{10} q$, where
$\epsilon$ is the difference between asymptotic and numerical
eigenvalues.  For evaluating the numerical eigenvalues, I leverage the
\texttt{eval\_lambda} tool bundled with release 6.0 and later
\footnote{Available for download at \url{https://github.com/rhdtownsend/gyre}}
of the open-source GYRE stellar oscillation code
\citep{Townsend:2013aa,Townsend:2018aa}. In brief, this tool solves the TEs using
the spectral matrix approach described by \citet{Townsend:2003ab},
implemented via a Sturm Sequence method
\citep[e.g.,][]{Barth:1967aa}. Initial eigenvalue brackets are
established using the asymptotic expressions themselves. At a given
$q$, the matrix dimension $N$ is determined dynamically by repeated
doubling until $\lambda$ converges to a fixed value in 64-bit
floating-point precision.

Each panel reveals a scaling $|\epsilon| \propto q^{-2}$ toward larger
$q$. This is the expected behavior of the asymptotic expressions, which
all claim an absolute accuracy of order $q^{-1}$. For the
gravito-inertial modes, the noise appearing for $\log_{10} q \gtrsim
3.5$ is due to the effects of rounding errors on the numerical
eigenvalues, rather than any issue with the asymptotic ones.

To explore how the eigenvalue expressions perform toward larger $m$
and $s$, Fig.~\ref{f:s-m-scan} plots $\log_{10} |\delta|$ as a
function of $m$, evaluated at fixed spin parameter $q=10^{3}$ for
selected meridional orders of each mode type. Here, $\delta \equiv
\epsilon/\lambda$ is the relative difference between asymptotic and
numerical eigenvalues. For the gravito-inertial modes, $|\delta|$
tends to decrease toward larger $s$, but increase toward larger
$|m|$. For the Rossby modes, the opposite trend is seen with respect
to $s$, while $|\delta|$ becomes independent of azimuthal order toward
large $|m|$. Finally, for the Kelvin modes $|\delta|$ decreases toward
larger $|m|$. The detailed reasons for these different behaviors lie
beyond the scope of this paper (since their elucidation would require
extending the asymptotic expressions to higher order in $q^{-1}$).


\section{Summary \& Discussion} \label{s:summary}

The principal results of this paper are the improved asymptotic
expressions for the eigenvalues of Laplace's tidal equations. For
gravito-inertial modes (equation~\ref{e:g-eigval}), the new expression
has a relative (absolute) accuracy of order $q^{-3}$ ($q^{-1}$), and
extends two orders in $q^{-1}$ further than the corresponding TO3 result.
For Rossby modes (equation~\ref{e:r-eigval}), the new expression has
an accuracy (both relative and absolute) of order $q^{-1}$, one order
in $q^{-1}$ further than T03. For Kelvin modes (equation~\ref{e:k-eigval}),
the new expression also has an accuracy of order $q^{-1}$ --- the same as
T03, but the latter did not formally establish the order of
correctness.

As one example application of these expressions, consider the
approximate formula
\begin{equation}
  P \equiv \frac{2\pi}{\omega} \approx \frac{\Pi_{0}}{\sqrt{\lambda}} \left( n + \frac{1}{2} \right)
\end{equation}
governing the co-rotating frame periods of low-frequency modes trapped
in the radiative zone between a convective core and a convective
surface layer. In this expression, which is derived from radial
asymptotic analysis within the TAR \citep[see, e.g.,][]{Bouabid:2013aa}, $n$ is the mode radial order and
$\Pi_{0}$ the asymptotic g-mode period spacing. Combining with
equation~(\ref{e:r-eigval}), I solve to obtain an explicit expression for
Rossby-mode periods,
\begin{equation}
  P \approx - \frac{(2s + 1) \Pi_{0}}{m} \left( n + \frac{1}{2} \right) - \pi \frac{1 + 2s(s + 1) \left[ 1 + 8 m^{2} - 4 s (s + 1) \right]}{4 (2 s + 1)^{2} m\, \Omega}
\end{equation}
Thus, for rotation sufficiently rapid that equation~(\ref{e:r-eigval})
provides a reasonable approximation, a sequence of Rossby modes with
the same $m$ and $s$ and consecutive $n$ should exhibit a uniform
period spacing within the co-rotating frame, equal to $(2s+1)/|m|$ times
the asymptotic g-mode period spacing. This result may prove useful in
analyzing recent identifications of these modes in $\gamma$ Doradus
stars \citep[e.g.,][]{Van-Reeth:2016aa,Li:2019aa}.

On a closing note, although this paper focuses on the eigenvalues of
Laplace's tidal equations, my analysis can also be used to construct
asymptotic expressions for the corresponding eigenfunctions. For
instance, combining
equations~(\ref{e:power-expand},\ref{e:r-Ht-0},\ref{e:r-Ht-1},\ref{e:r-c-1})
leads to an expression for Rossby-mode Hough functions,
\begin{dmath}
  \Ht = c_{0} \left\{
  \psi_{s} -
  \frac{(2s + 1)}{8m} \left[
    \frac{\sqrt{s (s-1) (s-2) (s-3)}}{4} \, \psi_{s-4} + 
    \frac{\sqrt{s (s-1)}}{s} \, \psi_{s-2} - 
    \frac{\sqrt{(s+1) (s+2)}}{s + 1} \, \psi_{s+2} -
    \frac{\sqrt{(s+1) (s+2) (s+3) (s+4)}}{4} \, \psi_{s+4}
    \right] \, q^{-1} + \mathcal{O}(q^{-2})
  \right\},
\end{dmath}
to accompany the eigenvalue expression~(\ref{e:r-eigval}). The
equivalent expression in TO3 (equation 32, ibid) included only the
first term in the braces.


\section*{Data Availability Statement}

The numerical data used to validate the asymptotic expressions are available on request from the author.


\section*{Acknowledgments}

I thank the anonymous referee for their helpful comments. This work
has been supported by NSF grants ACI-1663696, AST-1716436 and
PHY-1748958, and has made extensive use of NASA's Astrophysics Data
System Bibliographic Services.




\bibliographystyle{mnras}
\bibliography{laplace-eig}



\appendix

\section{Normalized Hermite Functions} \label{a:hermite}

The normalized Hermite functions $\psi_{j}$ ($j = 0,1,2\,\ldots$) are
defined in terms of the Hermite polynomials $H_{j}$ as
\begin{equation} \label{e:hermite-func}
  \psi_{j}(\sigma) = \frac{1}{\left( \sqrt{\pi} 2^{j} j! \right)^{1/2}} \, \exp \left( -\sigma^{2}/2 \right) \, H_{j}(\sigma).
\end{equation}
\citep[see, e.g., Section 18.2 of][]{Arfken:2013aa}. They are
orthonormal on the interval $[-\infty,\infty]$,
\begin{equation} \label{e:orthnorm}
  \left\langle \psi_{j}, \psi_{k} \right\rangle = \delta_{j,k},
\end{equation}
where
\begin{equation}
  \label{e:inner-prod}
  \left\langle f, g \right \rangle \equiv \int_{-\infty}^{\infty} f(\sigma)\, g(\sigma) \,\diff\sigma
\end{equation}
defines the inner product between the functions $f$ and $g$, and
$\delta_{j,k}$ is the Kronecker delta. As such, they form a complete
orthogonal basis for square-integrable real functions.

The normalized Hermite functions obey the identities
\begin{gather}
  \left( \nderiv{2}{}{\sigma} - \sigma^{2} \right) \psi_{j} = -(2j + 1) \, \psi_{j}, \\
  \sigma \deriv{\psi_{j}}{\sigma} = \frac{\sqrt{j(j-1)}}{2} \psi_{j-2} - 
    \frac{1}{2} \, \psi_{j} -
    \frac{\sqrt{(j + 1)(j + 2)}}{2} \, \psi_{j+2}, \\
  \sigma^{2} \psi_{j} =
    \frac{\sqrt{j (j -1)}}{2} \, \psi_{j-2} +
    \frac{2 j + 1}{2} \, \psi_{j} +
    \frac{\sqrt{(j + 1)(j + 2)}}{2} \, \psi_{j+2};
\end{gather}
these are used extensively in evaluating the inner products appearing
in Section~(\ref{s:asymp}).


\bsp	
\label{lastpage}
\end{document}